\documentclass{moriond}

\def\be{\begin{equation}}
\def\ee{\end{equation}}
\def\bea{\begin{eqnarray}}
\def\eea{\end{eqnarray}}

\newcommand{\Photo}{\includegraphics[height=35mm]{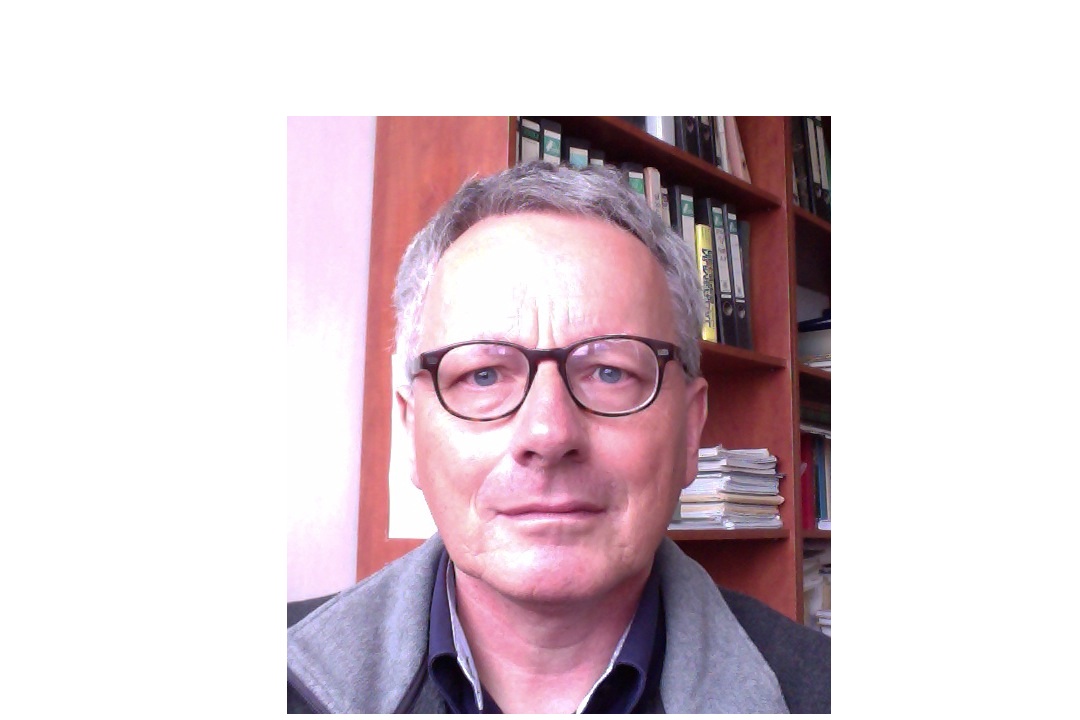}}

\begin{document}
\vspace*{4cm}
\title{GEOMETRICAL SCALING AND ITS BREAKING \\ 
IN HIGH ENERGY COLLISIONS~\footnote{Presented at the 49-th Rencontres de Moriond, 
March 22-29, 2014, La Thuille, Italy.}}

\author{ M. PRASZALOWICZ }

\address{M. Smoluchowski Institute of Physics, Jagiellonian University\\
4 Reymonta str., 30-059 Krakow, Poland}

\maketitle\abstracts{
We report on recent analyses of different pieces of data, which exhibit Geometrical Scaling (GS)
and its breaking. GS is a consequence of the existence of an intermediate energy scale, called
saturation momentum, and allows to relate data at different energies, of different systems
and also at different multiplicities and/or centralities.
}

In this talk we give a short overview of searches for the presence of Geometrical Scaling in  hadronic 
collisions. For details we refer the reader to the original publications. Let's 
start from the formula for the cross-section for inclusive gluon production~\cite{Gribov:1981kg} 
in a collision $1+2\rightarrow g +X$:
\begin{equation}
\frac{d\sigma}{dyd^{2}p_{\rm T}}=\frac{3\pi}{2p_{\rm T}^{2}}%
{\displaystyle\int}
d^{2}\vec{k}_{{\rm T}}\,\alpha_{{\rm s}}(k_{{\rm T}}^{2})\varphi_{1}%
(x_{1},\vec{k}_{{\rm T}}^{2})\varphi_{2}(x_{2},(\vec{k}-\vec{p}\,)_{{\rm T}%
}^{2}).\label{sigma_def}%
\end{equation}
Here $\varphi_{1,2}$ are unintegrated gluon densities
and $x_{1,2}$ are gluon momenta
fractions needed to produce a gluon of transverse momentum $p_{{\rm T}}$ and rapidity
$y$:
\begin{equation}
x_{1,2}=e^{\pm y}p_{{\rm T}}/\sqrt{s}\label{x12}\, .%
\end{equation}

Note that unintegrated gluon densities have dimension of area. 
This is at best seen from the very simple parametrization
propsed by Kharzeev and Levin~\cite{Kharzeev:2001gp} or by 
Golec-Biernat and W{\"u}sthoff~\cite{GolecBiernat:1998js} in the context of Deep
Inelastic Scattering (DIS):%
\begin{equation}
\varphi(k_{{\rm T}}^{2})=S_{\bot}\left\{
\begin{array}
[c]{rrr}%
1 & {\rm for} & k_{{\rm T}}^{2}<Q_{{\rm s}}^{2}\\
Q_{{\rm s}}^{2}/k_{{\rm T}}^{2} & {\rm for} & k_{{\rm T}}^{2}<p_{{\rm T}%
}^{2}%
\end{array}
\right.  \;{\rm or\ }\;\varphi(k_{{\rm T}}^{2})=S_{\bot}\frac{3}{4\pi^{2}%
}\frac{k_{{\rm T}}^{2}}{Q_{{\rm s}}^{2}}\exp\left(  -k_{{\rm T}}%
^{2}/Q_{{\rm s}}^{2}\right)  .\label{glue}%
\end{equation}
Here $S_{\bot}$ is the transverse size given by inelastic cross-section (or
its part) for the minimum bias inclusive multiplicity or in the case of DIS
$S_{\bot}=\sigma_{0}$ is the dipole-proton cross-section for large dipoles.
Another feature of the unintegrated glue (\ref{glue}) is the fact that
$\varphi$ depends on the ratio $k_{{\rm T}}^{2}/Q_{{\rm s}}^{2}(x)$ rather
than on $k_{{\rm T}}^{2}$ and $x$ separately. This is called Geometrical 
Scaling~\cite{Stasto:2000er} and has been for the first time proposed 
in the context of DIS. Here
\begin{equation}
Q_{{\rm s}}^{2}(x)=Q_{0}^{2}\left(  x/x_{0}\right)  ^{-\lambda}=Q_{0}%
^{2}(e^{\pm y}\, p_{{\rm T}}/W)^{-\lambda}\label{QspT}%
\end{equation}
is the saturation scale. Taking~\footnote{The precise value of $Q_{0}$
and $x_{0}$ is not important in the following. Only the value of exponent
$\lambda$ will be determined.}
 $x_{0}=10^{-3}$ we have 
$W=\sqrt{s}\times 10^{-3}$ in formula (\ref{QspT}).

Assuming particles $1$ and 2 to be identical and $y\sim0$ (central rapidity)
which corresponds to $x_{1}\simeq x_{2}$ (denoted in the following as $x$) and
suppressing $\alpha_{{\rm s}}$ we arrive at:%
\begin{equation}
\frac{d\sigma}{dyd^{2}p_{{\rm T}}}=S_{\bot}^{2}\mathcal{F}(\tau
)\qquad{\rm or}\qquad\frac{1}{S_{\bot}}\frac{dN}{dyd^{2}p_{{\rm T}}%
}=\mathcal{F}(\tau)\label{sigma_2}%
\end{equation}
where $\tau=p_{{\rm T}}^{2}/Q_{{\rm s}}^{2}(x)$ is scaling variable and
$dN/dy$ stands for multiplicity density. Eq.(\ref{sigma_2}) implies that
particle spectra at different energies should coincide if plotted in terms $\tau$.
In other words they exhibit GS~\cite{McLerran:2010ex}.

We can integrate now (\ref{sigma_2})
over $d^{2}p_{{\rm T}}$ using%
\[
dp_{{\rm T}}^{2}=\frac{2Q_{0}^{2}}{2+\lambda}\left(  W^{2}/Q_{0}^{2}\right)
^{\frac{\lambda}{2+\lambda}}\tau^{-\frac{\lambda}{2+\lambda}}d\tau
\]
arriving at%
\begin{equation}
\frac{dN}{dy}=S_{\bot}%
{\displaystyle\int}
\mathcal{F}(\tau)d^{2}p_{{\rm T}}=S_{\bot}\bar{Q}_{{\rm s}}^{2}\frac{2\pi
}{2+\lambda}%
{\displaystyle\int}
\mathcal{F}(\tau)\tau^{-\frac{\lambda}{2+\lambda}}d\tau=\frac{1}{\kappa
}S_{\bot}\,\bar{Q}_{{\rm s}}^{2}\label{dsdy}%
\end{equation}
where $1/\kappa$ is a universal, energy independent integral of $\mathcal{F}$, and
\begin{equation}
\bar{Q}_{{\rm s}}^{2}=Q_{0}^{2}\left(  W^{2}/Q_{0}^{2}\right)  ^{\frac
{\lambda}{2+\lambda}}\label{Qbars}%
\end{equation}
is an \emph{average} saturation scale, which can be thought of as a solution
of the equation%
\[
Q_{{\rm s}}^{2}(\bar{Q}_{{\rm s}}^{2}/W^2)=\bar{Q}_{{\rm s}}^{2}.
\]
It follows that%
\begin{equation}
\bar{Q}_{{\rm s}}^{2}=\frac{\kappa}{S_{\bot}}\frac{dN}{dy}.\label{QsdNdy}%
\end{equation}
Equation (\ref{QsdNdy}) means that the average saturation scale is
proportional to the gluon density per unit transverse area. One should
keep in mind the distinction between saturation scales (\ref{QspT}) and (\ref{QsdNdy}),
since they are interchangeably used in the literature. The theory behind
gluon saturation (for a review see Refs.~[6,7]\nocite{Mueller:2001fv,McLerran:2010ub} 
and references therein)  is Color Glass Condensate \cite{sat1,sat2,MLV}.

The existence of GS in pp collisions as given by eq.(\ref{sigma_2}) has been
indeed observed in the data~\cite{McLerran:2010ex} and reported 
at Moriond 2012~\cite{Praszalowicz:2012ab}. An efficient way to
study GS is to form ratios $R_{W_{1},W_{2}}(\tau)=\left.  dN/dydp_{{\rm T}%
}\right\vert _{W_{1}}(\tau)/\left.  dN/dydp_{{\rm T}}\right\vert _{W_{2}%
}(\tau)$ which, according to (\ref{sigma_2}) should be equal $1$ over wide
range of $\tau$~\cite{Praszalowicz:2011rm}. This requirement allows to 
find the optimal value of
$\lambda$ which in the case of the LHC data is equal to 0.27, which is a bit smaller
than in DIS~\cite{Praszalowicz:2012zh}.  It has been shown~\cite{Praszalowicz:2012zh} 
that in DIS GS extends up to rather large $x_{\rm max} \approx 0.08$.

\begin{figure}[h]
\centering
\includegraphics[width=7.0cm,angle=0]{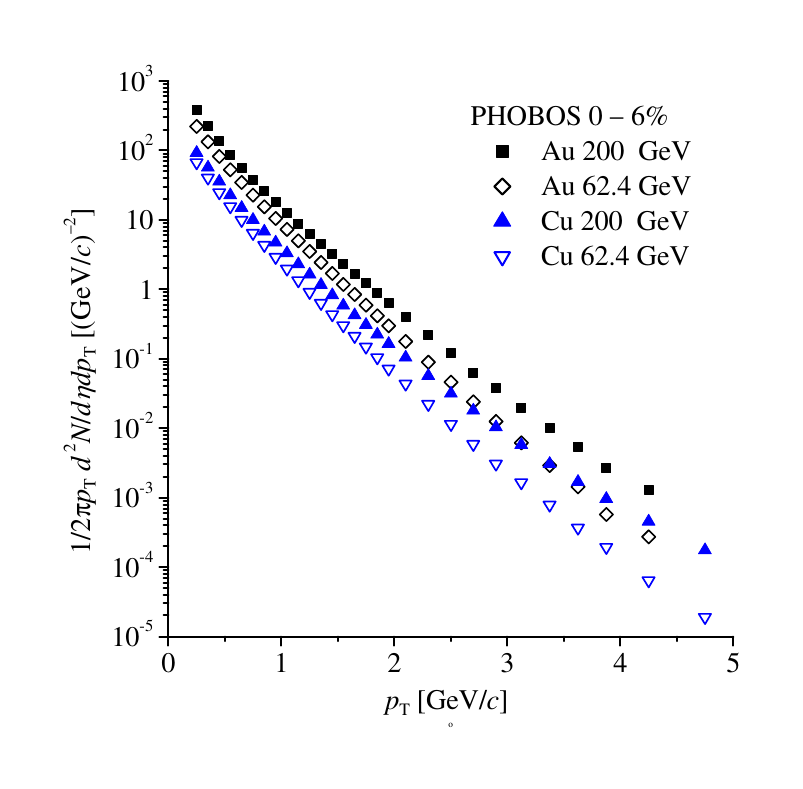}
\includegraphics[width=7.0cm,angle=0]{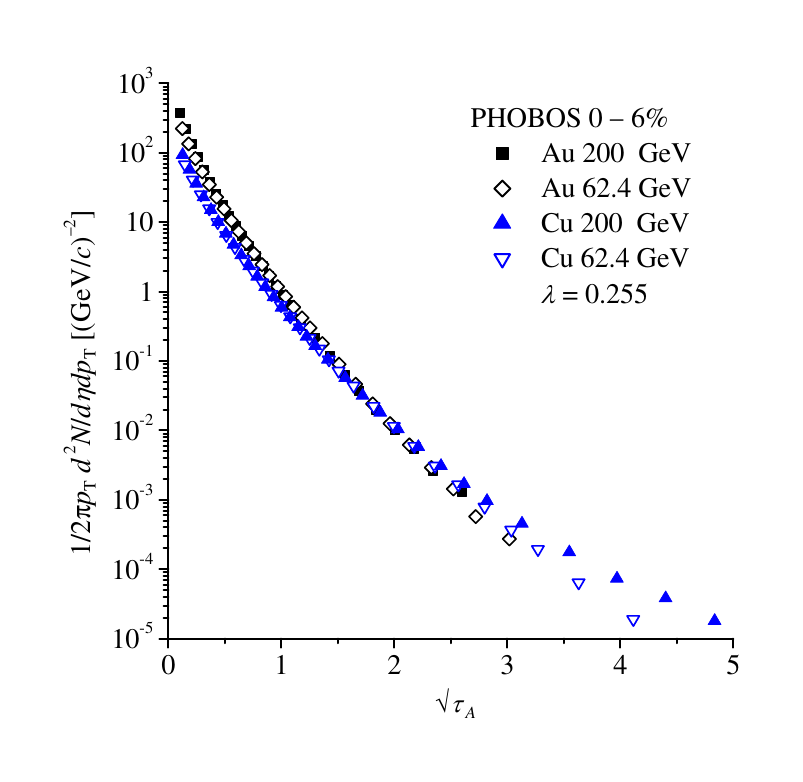}\caption{Transverse
momentum spectra measured by PHOBOS in Au-Au and Cu-Cu most central collisions
as functions of $p_{\mathrm{T}}$ (right) and $\sqrt{\tau_{A}}$ (left)
(from Ref.[~{\protect\cite{Praszalowicz:2011rm}]})}%
\label{PHOBOS}%
\end{figure}

Surprisingly GS scaling works also for the $p_{{\rm T}}$ spectra in heavy ion
collisions at RHIC energies~\cite{Praszalowicz:2011rm}. In the case of heavy 
ions the saturation momentum
scales as $Q_{A\,{\rm s}}^{2}=A^{1/3}Q_{{\rm s}}^{2}$ and the scaling
variable is therefore $\tau_{A}=p_{{\rm T}}^{2}/Q_{A\,{\rm s}}^{2} $. This
is illustrated in Fig.~\ref{PHOBOS} where charged particles spectra in AuAu and CaCa 
collisions as
measured by PHOBOS are plotted in terms of $p_{{\rm T}}$ and $\sqrt{\tau_{A}%
}.$ Recently GS for the photons produced in different systems (AA, dA and pp), at different energies 
and at different centralties ({\em i.e.} at different $S_{\bot}$) has been reported~\cite{Klein-Bosing:2014uaa}.

For $y>0$ two Bjorken $x$'s (\ref{x12}) can be quite different: $x_{1}>x_{2}$.
Therefore by increasing $y$ one can eventually reach $x_{1}>x_{\mathrm{max}}$
and violation of GS is expected. To show this~\cite{Praszalowicz:2013uu}  
we have used pp data from NA61/SHINE
experiment \cite{NA61} which measured particle spectra at different rapidities
$y=0.1-3.5$ and at 5 scattering energies $W_{1,\ldots,5}%
=17.28,\;12.36,\;8.77,\;7.75$, and $6.28$ GeV.


\begin{figure}[h]
\centering
\includegraphics[width=7.2cm,angle=0]{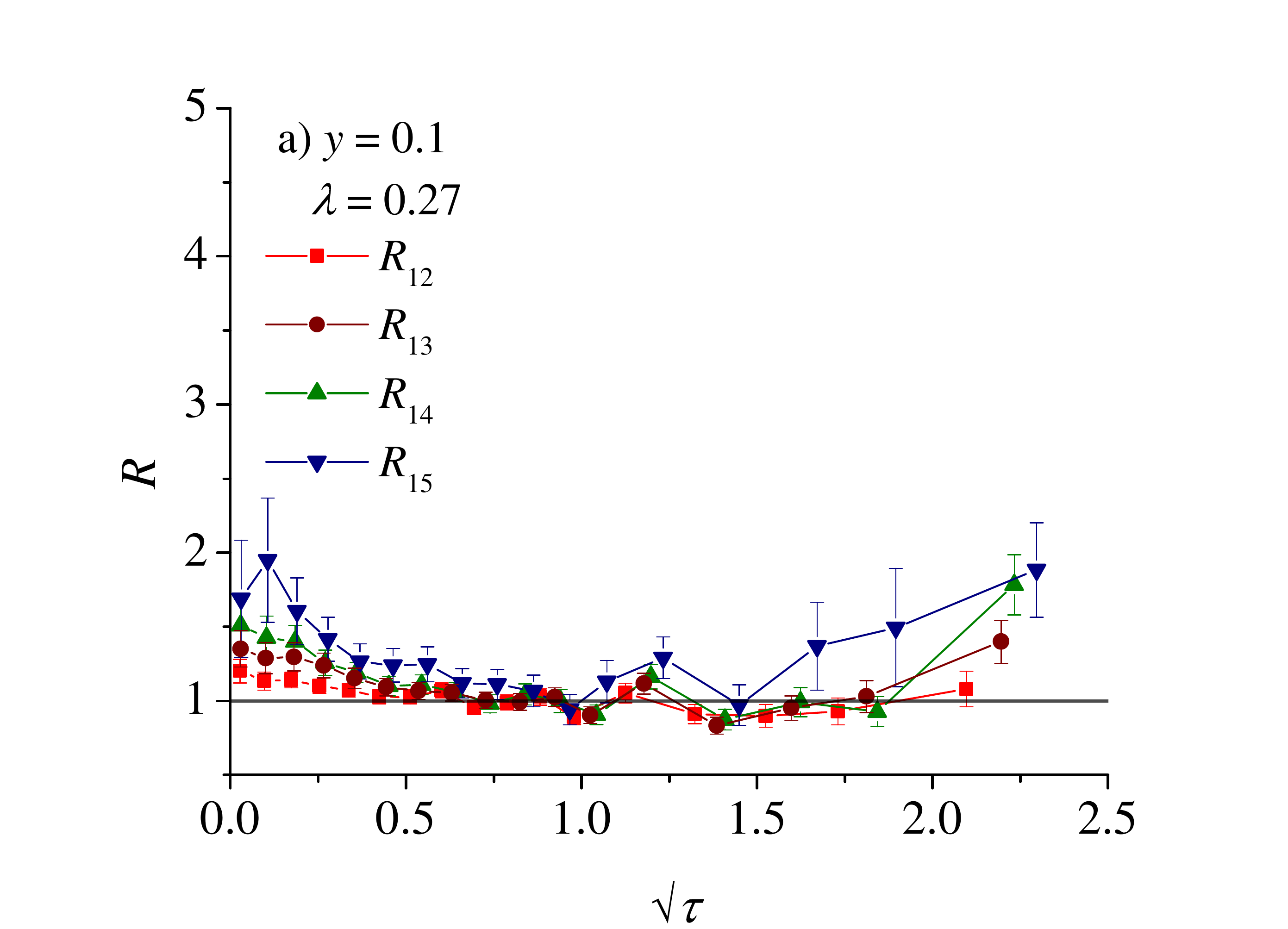}
\includegraphics[width=7.2cm,angle=0]{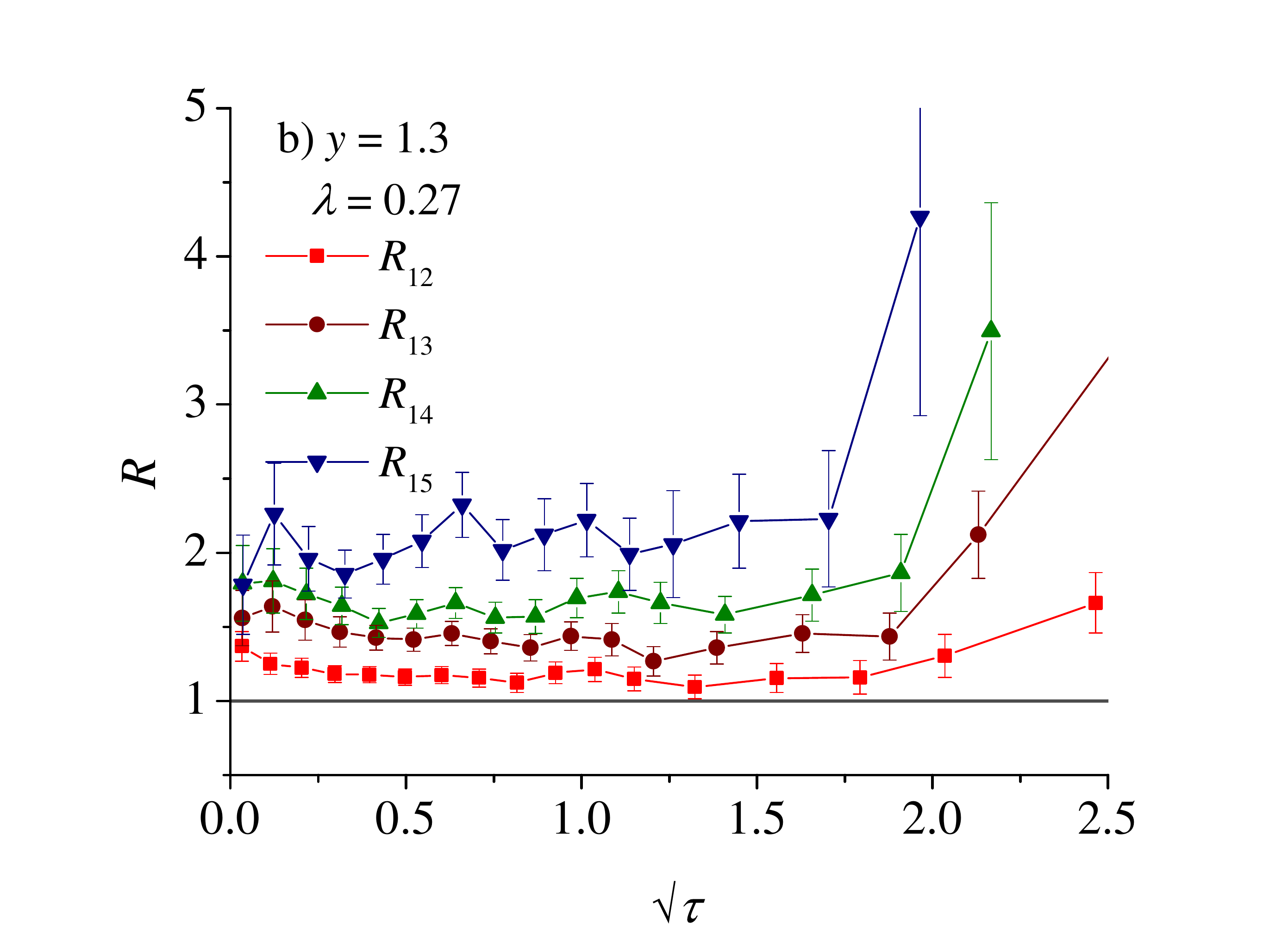}\caption{Ratios $R_{1k}$
as functions of $\sqrt{\tau}$ for $\lambda=0.27$ and for different rapidities
a) $y=0.7$ and b) $y=1.3$. With increase of rapidity, gradual closure of the
GS region can be seen (from Ref.[~{\protect\cite{Praszalowicz:2013uu}]}).}%
\label{ys}%
\end{figure}

In Fig.~\ref{ys}.a we plot ratios $R_{1i}=R_{W_{1},W_{i}}$ for
$\pi^{-}$ spectra in rapidity region $y=0.1$ for $\lambda=0.27$. Here the GS
window extends down to the smallest energy because $x_{\mathrm{max}}$ is as
large as 0.08. Nevertheless one can see that the quality of GS is the worst
for the smallest energy $W_{5}$. By increasing $y$ some points fall outside
the GS region due to the fact that $x_{1}\geq x_{\mathrm{max}}$, and finally
for $y\geq1.7$ geometrical scaling is no longer seen. This is shown in
Fig.~\ref{ys}.b.

In a situation where two (or more) external energy scales are present, like
$p_{\mathrm{T}}$ and particle mass $m$ (for identified particles), one can
form two independent ratios with $Q_{\mathrm{s}}$ what implies violation or at
least modification of GS. We have argued
that in the case of identified particles GS is still present~\cite{Praszalowicz:2013fsa} 
if another
scaling variable is used in which $p_{\mathrm{T}}$ is replaced by $\tilde
{m}_{\mathrm{T}} =m_{\mathrm{T}} - m = \sqrt{m^{2}_{\mathrm{T}} +p^{2}%
_{\mathrm{T}}}-m $. This scaling
variable is connected with the fact that  accurate fits are obtained by means
of Tsallis-like parametrization \cite{Tsallis,Rybczynski:2012vj,Cleymans:2013rfq} 
where particle multiplicity
distribution takes the following form (see \emph{e.g.}~Ref.[21]
\nocite{Chatrchyan:2012qb}):
\begin{equation}
\frac{1}{p_{{\rm T}}}\frac{d^{2}N}{dydp_{{\rm T}}}=C\frac{dN}{dy}\left[
1+\frac{m_{\mathrm{T}}-m}{n\,T}\right]  ^{-n}. \label{Tsallis}%
\end{equation}
Coefficient $C$  ensures proper normalization of
(\ref{Tsallis}). Here $n$ and $T$ are free fit parameters that depend on
particle species. Formula (\ref{Tsallis}) admits GS solution~\cite{Praszalowicz:2013fsa}, 
provided that
$n$ is a constant (with possible corrections that would allow for the energy
dependence of $n$ seen in the data) and $T \sim\bar{Q}_{\mathrm{s}}$ of
eq.(\ref{Qbars}) which has a power-like energy dependence.

In summary we can say that by now the existence of the saturation scale is
undoubtedly well established. Geometrical Scaling follows as a natural consequence.
One can use GS to relate different pieces of data with an accuracy much higher than
originally expected. New results from the LHC at higher energies will be  important
for further studies of the details GS and of the underlying theory of dense gluonic system.

\section*{Acknowledgemens}

The author would like to thank the organizers E. Aug{\'e} and B. Pietrzyk. 
This work was supported by the Polish NCN grant 2011/01/B/ST2/00492.

\end{document}